\documentclass[aps,prl,secnumarabic,nobalancelastpage,amsmath,amssymb,twocolumn]{revtex4}

\usepackage{graphics}
\usepackage{graphicx}

\begin{document}
\title{Observation of a Free-Shercliff-Layer Instability in Cylindrical Geometry}
\author{Austin H. Roach}
\email{aroach@pppl.gov}
\author{Erik J. Spence}
\author{Christophe Gissinger}
\author{Eric M. Edlund}
\author{Peter Sloboda}
\author{Jeremy Goodman}
\author{Hantao Ji}
\email{hji@pppl.gov}
\affiliation{Center for Magnetic Self-Organization in Laboratory and Astrophysical Plasmas and Princeton Plasma Physics Laboratory, P.O. Box 451 Princeton, New Jersey 08543, USA}
\date{\today}

\begin{abstract}
We report on observations of a free-Shercliff-layer instability in a
Taylor-Couette experiment using a liquid metal over a wide range of
Reynolds numbers, $Re\sim 10^3-10^6$. The free Shercliff layer is
formed by imposing a sufficiently strong axial magnetic field across a
pair of differentially rotating axial endcap rings. This layer is
destabilized by a hydrodynamic Kelvin-Helmholtz-type instability,
characterized by velocity fluctuations in the $r-\theta$ plane. The
instability appears with an Elsasser number above unity, and saturates
with an azimuthal mode number $m$ which increases with the Elsasser
number. Measurements of the structure agree well with 2D global linear
mode analyses and 3D global nonlinear simulations. These observations
have implications for a range of rotating MHD systems in which
similar shear layers may be produced.
\end{abstract}

\maketitle

The destabilization of rotating sheared flows by an applied magnetic
field in magnetohydrodynamics (MHD) is a topic with astrophysical
and geophysical implications, and has been the subject of a
number of experimental and theoretical efforts. Such destabilization
can be caused by the magnetorotational instability (MRI), in which a
magnetic field of sufficient amplitude can destabilize Rayleigh-stable
rotating sheared flows~\cite{balbus.apj.1991}. In this Letter, we
report the observation of an instability which, like the MRI, appears
in a sheared rotating fluid when a magnetic field is applied. But
rather than playing a role in the dynamics of the instability, as in
the case of the MRI, the magnetic field here acts to establish free
shear layers which extend from axial boundaries and which are subject
to a hydrodynamic instability.

Hartmann and Shercliff laid the groundwork in understanding the effect
of magnetic fields on shear layers in conducting fluids. Hartmann
studied boundary layers normal to an external applied
field~\cite{hartmann.mfm.1937}, and Shercliff extended his analysis
to include boundary layers parallel to the applied
field~\cite{shercliff.mpcps.1953}. Free Shercliff layers can be
established in rotating MHD systems when the line-tying force of
an axial magnetic field extends a discontinuity in angular velocity at
an axial boundary into the bulk of the fluid. These shear layers are
similar to the Stewartson layers that extend from discontinuous
shearing boundaries in rapidly rotating
systems~\cite{stewartson.jfm.1957}, but for the free Shercliff layer
discussed here, it is the magnetic field tension rather than the
Coriolis force that leads to equalization of the angular velocity in
the axial direction.

Free Shercliff layers were first realized experimentally by Lehnert in
a cylindrical apparatus with a free surface at the top and a rotating
ring at the bottom axial boundary~\cite{lehnert.prs.1955}. Lehnert
observed the formation of vortices at the location of the shear
layers, though he attributed their formation to discontinuities in the
free surface at the shear layer location rather than to the shear
itself. These layers were then described analytically by
Stewartson~\cite{stewartson.qjmam.1957} and
Braginskii~\cite{braginskii.jetp.1960}.  The formation of free
Shercliff layers in a cylindrical Taylor-Couette device has been
predicted computationally~\cite{liu.pre.2008}, but these simulations
were axisymmetric and thus incapable of evaluating the stability of
these shear layers to nonaxisymmetric perturbations.

Both free Shercliff layers and Stewartson layers can be present at the
tangent cylinder of spherical Couette systems. The Kelvin-Helmholtz
destabilization of these layers has been studied extensively through
computation~\cite{hollerbach.prsla.2001, hollerbach.tcfd.2004,
wei.pre.2008}. Stewartson layers have been observed experimentally in
spherical and cylindrical geometry and are found to be unstable to
nonaxisymmetric modes, which is consistent with
simulations~\cite{hollerbach.tcfd.2004, schaeffer.pof.2005,
hide.jfm.1967}.

\begin{figure}
\centering
\includegraphics[width=0.95\columnwidth]{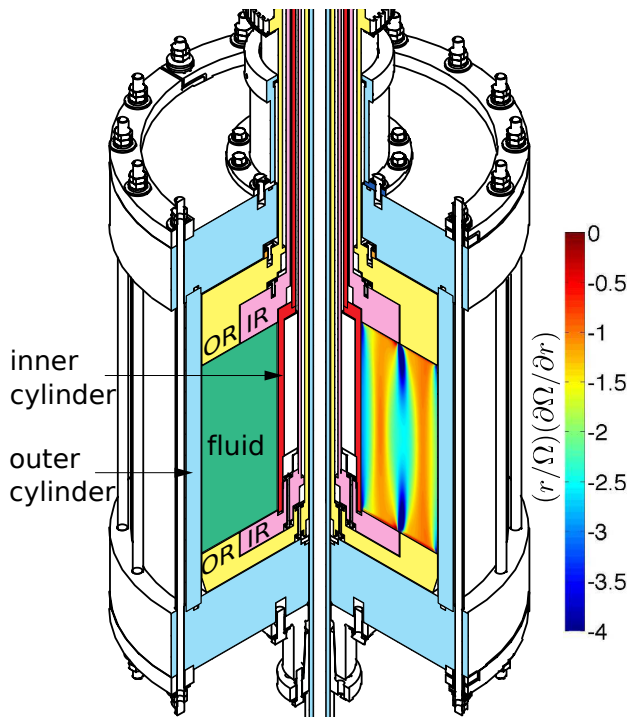}
\caption{Diagram of Princeton MRI experiment. Each endcap is split
into an inner ring (IR) and an outer ring (OR). Differential rotation
of these rings produces a discontinuity in the angular velocity at the
axial boundary. Overlaid on the right half of the figure is a plot of
the shear $(r/\Omega)(\partial \Omega/\partial r)$ from a nonlinear
MHD simulation with differential rotation between the endcap rings and
a strong axial magnetic field ~\cite{gissinger.pof.2012}. The
free Shercliff layers are the regions of strong negative shear
extending from the interface between the rings.}
\label{fig:experiment}
\end{figure}

The Princeton MRI experiment is a Taylor-Couette apparatus consisting
of two coaxial stainless steel cylinders as shown in
Fig.~\ref{fig:experiment}. The gap between the cylinders is filled
with a GaInSn eutectic alloy which is liquid at room
temperature. Differential rotation of the cylinders sets up a sheared
rotation profile in the fluid. If the cylinders were infinitely long,
the fluid between the cylinders would assume an angular velocity
$\Omega$ at a radius $r$ matching the ideal Couette solution in steady
state, $\Omega(r) = a + b/r^2$. The constants $a$ and $b$ are found by
matching the solution to the imposed rotation rates at the inner and
outer cylinder boundaries. In conventional Taylor-Couette devices, the
endcaps are typically corotated either with the inner or outer
cylinder. This produces strong secondary circulation and angular
momentum transport to the axial boundaries, resulting in a deviation
from the ideal rotation profile~\cite{kageyama.jpsj.2004}. A novelty
of this apparatus is the configuration of the axial endcaps, each of
which is split into two differentially-rotatable acrylic rings, giving
four independent rotation rates: those of the inner cylinder, outer
cylinder, inner rings, and outer rings. In previous experiments using
water as the working fluid, this configuration was very effective at
reducing the influence of the axial boundaries, allowing the
generation of quiescent flows in the bulk of the fluid with Reynolds
numbers $Re=\Omega_1 r_1 (r_2-r_1)/\nu$ above
$10^6$~\cite{ji.nature.2006}.  The experimental parameters are shown
in Table~\ref{table:parameters}.
 
Fluid velocities are measured with an ultrasound Doppler velocimetry
(UDV) system\cite{takeda.ned.1991, signalprocessing}. Ultrasonic
transducers are mounted on the outer cylinder at the midplane of the
experiment. A transducer aimed radially and others aimed tangential to
the inner cylinder allow determination of the radial and azimuthal
velocity components. Two tangential transducers aimed identically but
separated azimuthally by $90^{\circ}$ provide information about
azimuthal mode structure.

A set of six solenoidal coils applies an axial magnetic field to the
rotating fluid. Fields below 800 Gauss can be applied indefinitely,
while the application time for higher fields is limited by the
resistive heating of the coils. An array of 72 magnetic pickup coils
placed beyond the outer cylinder measures $\partial B_{r}/\partial t$.

\begin{table}
\caption{Parameters of the apparatus~\cite{schartman.rsi.2009} and
liquid metal working fluid~\cite{morley.rsi.2008}.}
\centering
\begin{tabular}{cccc}
\hline\hline
Parameter & symbol & value & units \\
\hline
Height & $h$ & 27.9 & cm \\
Inner cylinder radius & $r_{1}$ & 7.06 & cm \\
Outer cylinder radius & $r_{2}$ & 20.3 & cm \\
Density & $\rho$ & 6.36 & g/cm$^3$ \\
Kinematic viscosity & $\nu$ & $2.98 \times 10^{-3}$ & cm$^2$/s \\
Magnetic diffusivity & $\eta$ & $2.57 \times 10^{3}$ & cm$^2$/s \\
Inner cylinder rotation rate & $\Omega_1$ & 0.25 - 800 & rpm \\
Axial magnetic field & $B$ & 0 - 4500 & Gauss \\
\hline
\end{tabular}
\label{table:parameters}
\end{table}

Experiments were run using both Rayleigh-stable and -unstable flow
states. The Rayleigh-stable states had component rotation speeds in
the ratio [1.0, 0.55, 0.1325, 0.1325] for the inner cylinder, inner
ring, outer ring, and outer cylinder, respectively. The ideal Couette
solution for these inner and outer cylinder speeds satisfies
Rayleigh's stability criterion that the specific angular momentum
increase with radius: $\partial(r^2\Omega)/\partial r > 0$. The ring
speeds were chosen empirically to generate an azimuthal rotation
profile in the hydrodynamic case that closely matches the ideal
Couette profile at the midplane. The Rayleigh-unstable states were
generated using component speeds in the ratio [1.0, 1.0, 0, 0]. These
flows violate Rayleigh's criterion and exhibit large velocity
fluctuations in the absence of a magnetic field.

A single run of this experiment starts with an acceleration phase of
two minutes, during which the sheared azimuthal flow develops. The
axial magnetic field is then applied, initially resulting in the
damping of hydrodynamic fluctuations. If the magnetic field is strong
enough to satisfy the requirement that the Elsasser number $\Lambda =
B^2/4\pi\rho\eta\Delta\Omega > 1$, where $\Delta\Omega$ is the
difference between the inner- and outer-ring rotation rates, the
instability grows up as a large-scale coherent mode. It manifests
itself as a fluctuation in the radial velocity and azimuthal velocity,
where significant perturbations of more than 10\% of the inner
cylinder speed are observed. An ultrasonic transducer inserted on a
probe and aimed axially at an endcap did not measure axial velocity
fluctuations when the instability was excited, suggesting that the
flow due to the instability is mainly in the $r-\theta$
plane. Correlated magnetic fluctuations are observed at the highest
rotation rates and applied fields. The instability develops on both
the Rayleigh-stable and -unstable backgrounds, and typical mode
rotation rates exceeds the outer cylinder rotation rate $\Omega_2$ by
$\sim 0.1(\Omega_1-\Omega_2)$.

\begin{figure}
\centering
\includegraphics[width=0.99\columnwidth]{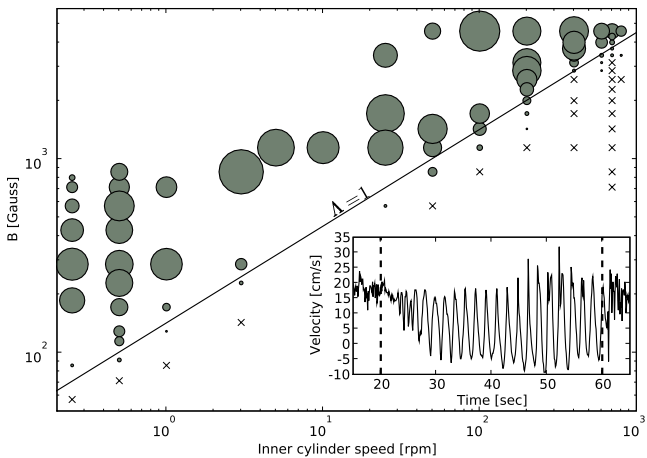}
\caption{Stability diagram for the Rayleigh-unstable background flow
state. The area of the circles is proportional to the power in the
dominant Fourier harmonic measured by a tangential transducer at
$r=19.2\mathrm{cm}$, normalized to the square of the inner cylinder
speed. The `x's indicate stability. The stability diagram for the case
starting from a Rayleigh-stable background is similar, but was
measured over a smaller range of speeds. The inset plot shows a sample
time trace of the velocity measured at one point in the flow, with the
magnetic field applied in the region between the dashed vertical
lines. ($\Omega_1=$200 rpm, $B=$2900 G.)}
\label{fig:stabilityplot}
\end{figure}

The instability was observed over a range of more than 3 orders of
magnitude in rotation rate in the Rayleigh-unstable configuration, as
shown in Fig.~\ref{fig:stabilityplot}, with
$Re=820-2.6\times10^{6}$. The instability is present even with a
magnetic Reynolds number $Rm=\Omega_1 r_1(r_2-r_1)/\eta \sim 10^{-3}$,
indicating an inductionless mechanism in which induced magnetic fields
are dynamically unimportant.

\begin{figure}
\centering
\begin{tabular}{cc}
\includegraphics[width=0.5\columnwidth]{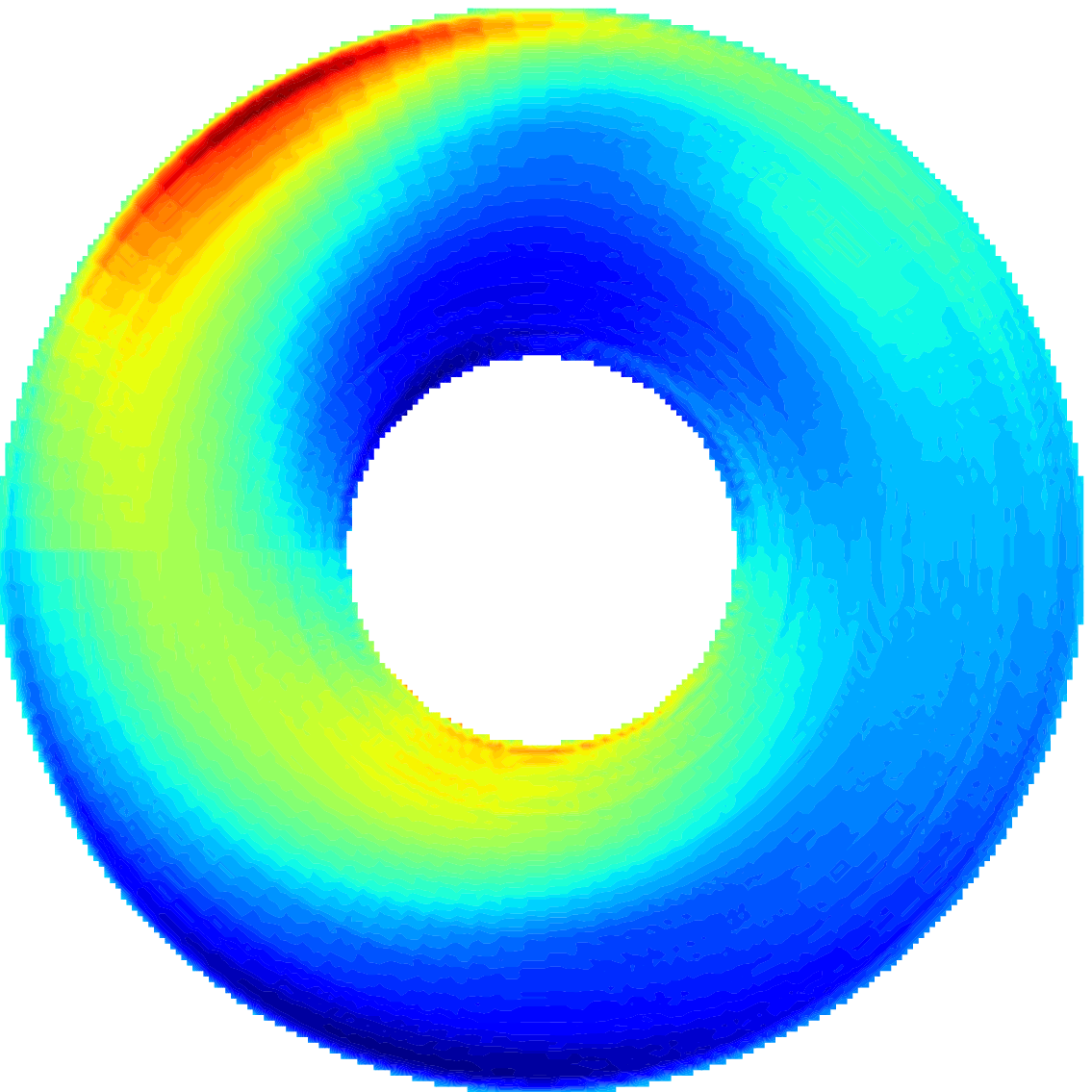} &
\includegraphics[width=0.5\columnwidth]{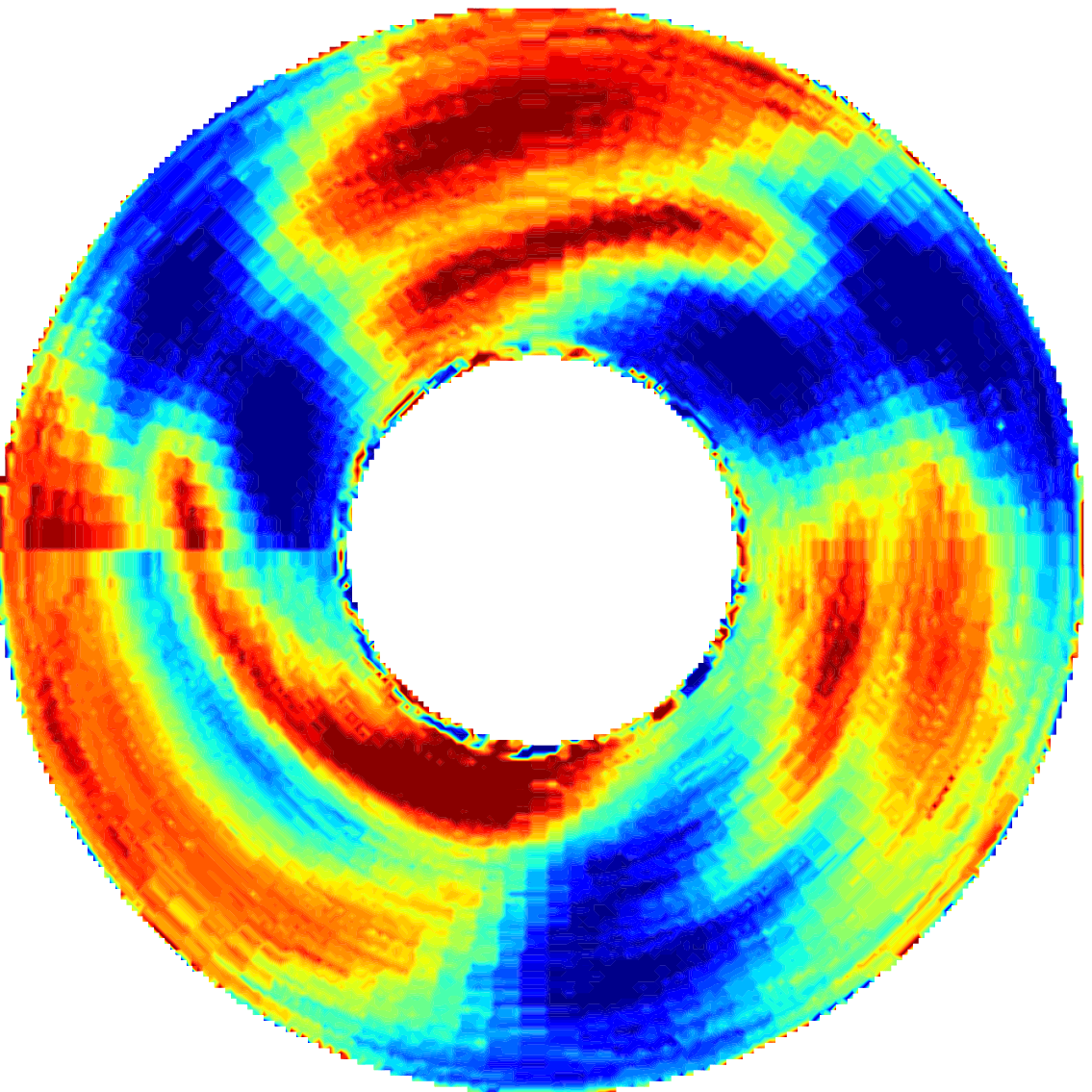} \\
\includegraphics[width=0.5\columnwidth]{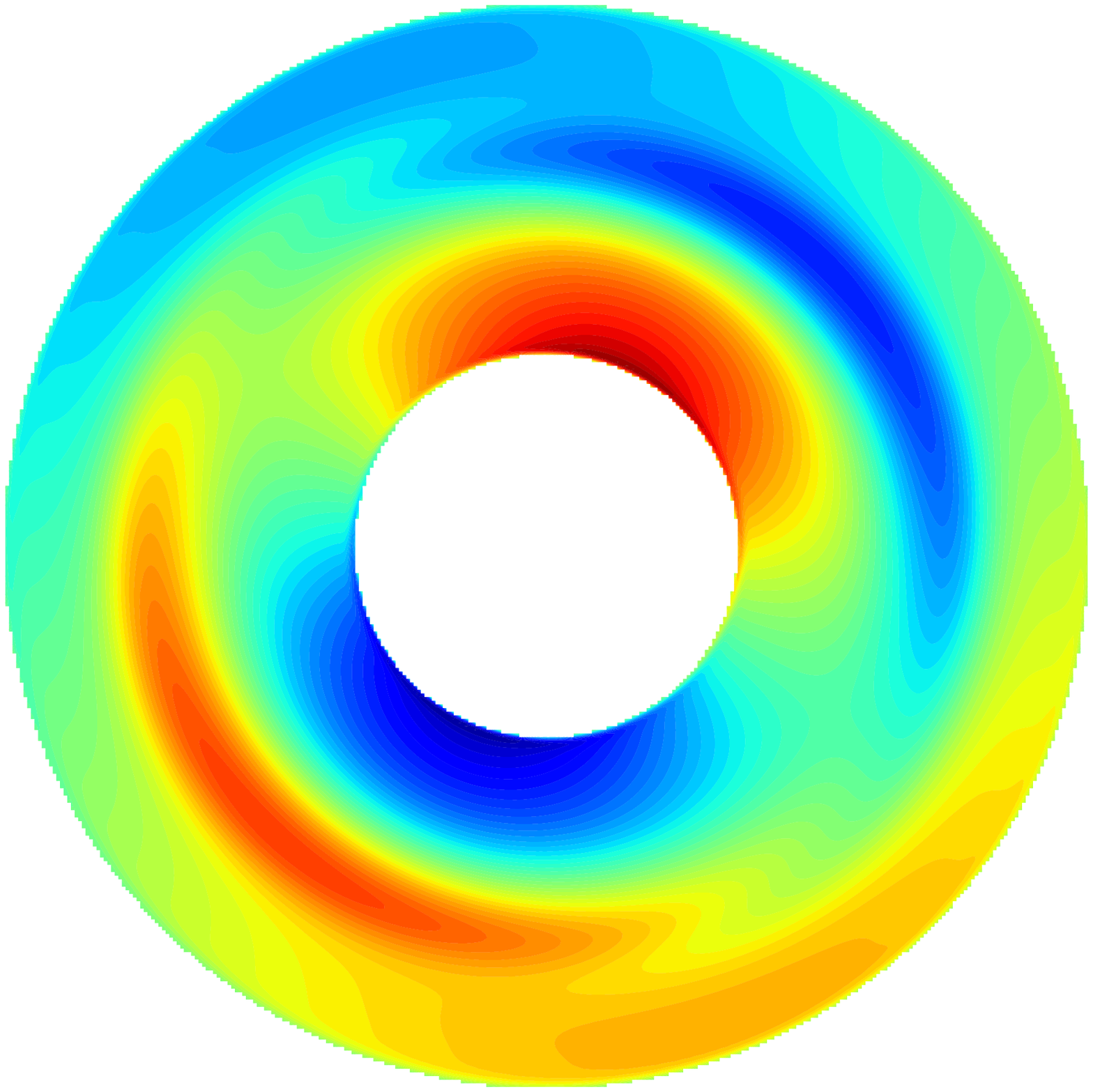} &
\includegraphics[width=0.5\columnwidth]{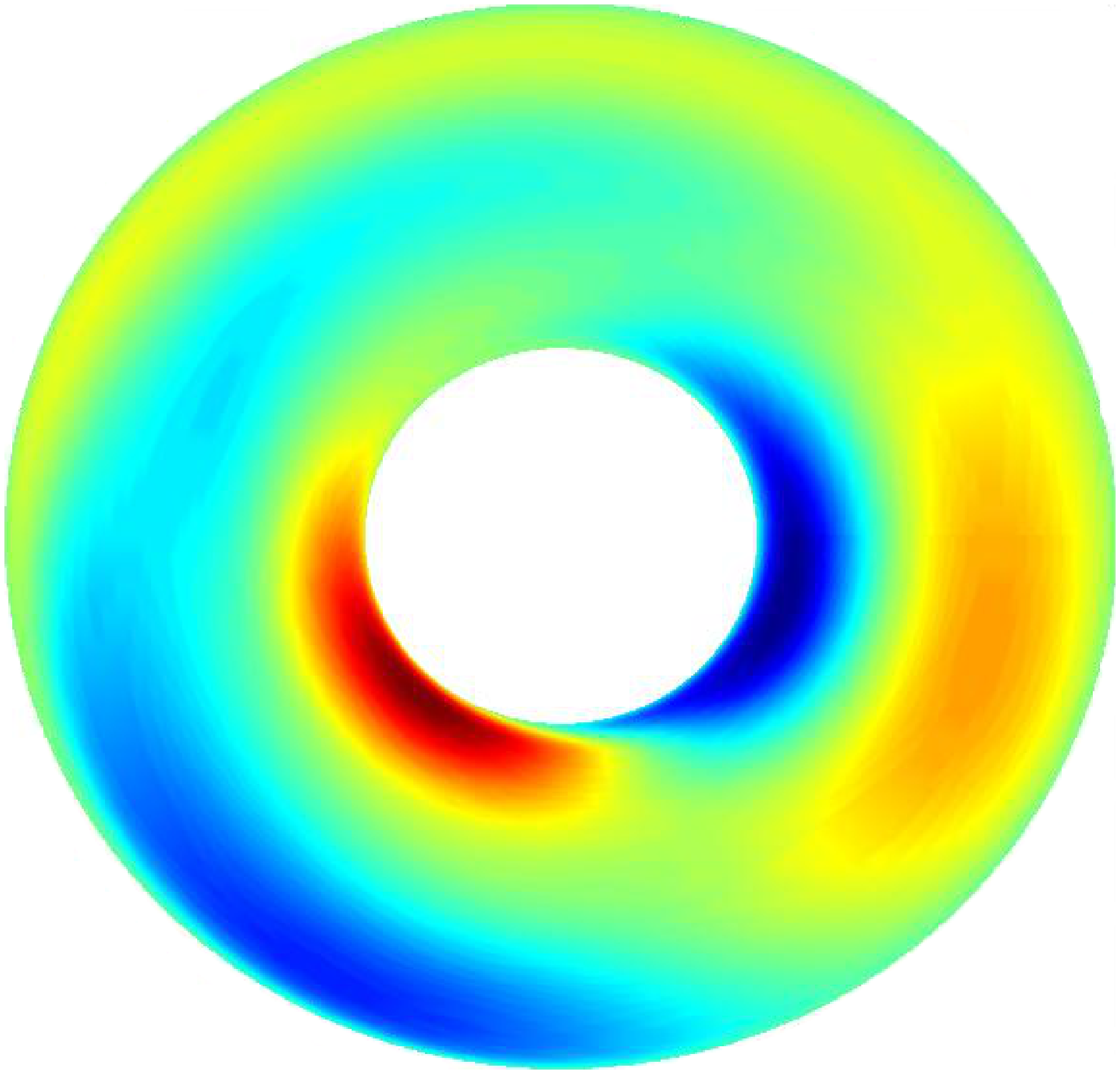} 

\end{tabular}
\caption{Comparison of measured unstable mode with results from
simulation. All are contour plots of azimuthal velocity at the
midplane with the $m=0$ contribution subtracted. Red indicates
positive velocity (counterclockwise), and blue indicates negative
velocity (clockwise). Upper left: Experimental measurement with
$\Lambda=1.4$, reconstructed from projecting the time behavior of the
azimuthal velocity as measured by one UDV transducer onto the
$r-\theta$ plane. Upper right: Experimental measurement with
$\Lambda=50$. Lower left: Growing $m=1$ mode produced by a
hydrodynamic linear stability analysis of an axially independent shear
layer. Lower right: Unstable mode from nonlinear MHD calculation with
$Re=4000$, $Rm=10$, and $\Lambda=1$.}
\label{fig:spiralmode}
\end{figure}

For $\Lambda$ of order one, the primary azimuthal mode number at
saturation is $m=1$, with phase-locked higher-order mode numbers
typically present at a smaller amplitude. The measured mode structure
is shown in Fig.~\ref{fig:spiralmode}. It is common for an $m=2$
mode to grow up before an $m=1$ dominates at
saturation. High-$\Lambda$ scenarios at very slow rotation rates show
that $m$ at saturation increases as $\Lambda$ increases. Primary mode
numbers up to $m=5$ have been observed with $\Lambda=127$ at a
rotation rate of 0.25 rpm.

The necessity of shear at the axial boundary has been verified
experimentally. Experiments were performed with the components
rotating in the standard Rayleigh-stable configuration, but with a
number of different inner ring speeds. The critical magnetic field for
instability varied with the differential rotation between the endcap
rings as expected. When the inner rings and outer rings corotated,
the instability was not observed.

\begin{figure}
\centering
\includegraphics[width=0.95\columnwidth]{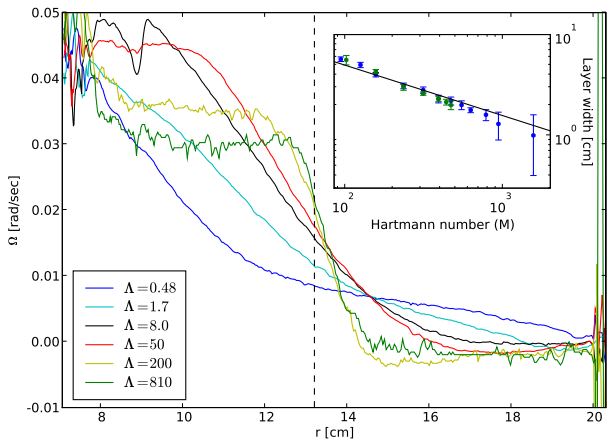}
\caption{Angular velocity versus radius at the midplane for several
values of applied fields at a rotation rate of 0.5 rpm in the
Rayleigh-unstable configuration. The dashed line indicates the radial
position of the split between the axial endcaps. The inset plot shows
measurements of the shear layer thickness from experiments at 0.25 rpm
(green) and 0.5 rpm (blue) versus the Hartmann number
$M$. $M=55\sqrt{\Lambda}$ at 0.5 rpm, and $M=39\sqrt{\Lambda}$ at 0.25
rpm. The solid line indicates the expected Shercliff layer width
scaling $\delta \sim d/\sqrt{M}$, where the constant $d=62\mathrm{cm}$ has been chosen to match the data.}
\label{fig:shearlayer}
\end{figure}

The free shear layer has been measured experimentally at low $Re$ and
high $\Lambda$ where it penetrates to the midplane of the experiment
as shown in Fig.~\ref{fig:shearlayer}. The width of the layer
measured at a time just before the onset of instability is consistent
with the expected width scaling for a Shercliff layer $\delta \sim
1/\sqrt{M}$, where the Hartmann number $M=Bl/\sqrt{4\pi\rho\eta\nu}$
and $l=r_2 - r_1$ is a characteristic length. The onset of the
instability is associated with a decrease in the mean shear in this
layer.

Nonlinear numerical MHD simulations have been performed with the
HERACLES code~\cite{gonzalez.aa.2007}, modified to include finite
viscosity and resistivity~\cite{gissinger.pof.2012}. The simulations
were performed in the experimental geometry with a 200x64x400 grid in
$\hat{r}$, $\hat{\theta}$, and $\hat{z}$, with $Re=4000$ and a range
of $Rm$ and $M$. These simulations show the formation of the free
Shercliff layer extending from the discontinuity at the axial
boundaries, as shown in Fig.~\ref{fig:experiment}. The axial length
of the shear layer scales with $\sqrt{\Lambda}$, which seems to arise
from a competition of magnetic forces, which act to extend the shear
layer into the fluid, and poloidal circulation generated by the axial
boundaries, which acts to disrupt the free shear layer. The
simulations also produce an instability requiring $\Lambda>1$ for
onset and suggest that a minimum penetration depth of the shear layer
is required for development of the instability. Like the experimental
observations, the unstable modes exhibit a spiral structure, and a
cascade is observed from higher azimuthal mode number during the
growth phase of the instability to a dominant $m=1$ at saturation.

A global linear stability analysis was performed to investigate
unstable modes in the experimental geometry. The analysis found
eigenvalues of the linearized nonideal MHD equations discretized
across 2048 grid cells in the radial direction, assuming sinusoidal
azimuthal dependence with a specified mode number and no axial
dependence. Unstable hydrodynamic solutions were sought for realistic
fluid parameters and for a zeroth order, background rotation profile
consisting of a free shear layer represented by a hyperbolic tangent
centered between the inner and outer cylinders. Angular velocity
profiles with a sufficiently narrow shear layer were found to be
hydrodynamically unstable to nonaxisymmetric Kelvin-Helmholtz modes
with a similar structure to those observed experimentally. The
most unstable mode number increases with decreasing shear layer width,
similar to the experimental observations of the saturated states.

The results presented here describe a minimum magnetic field required
for onset of the instability. Simulations have shown that a
sufficiently strong magnetic field will restabilize this instability,
similar to the simulation results in spherical
geometry~\cite{hollerbach.prsla.2001}. Experimentally, the decreasing
saturated amplitude with increasing field at small rotation rates,
shown in Fig.~\ref{fig:stabilityplot}, suggests that this critical
field strength is being approached. But the limits on controllable
slow rotation and on the availability of strong magnetic fields
precluded verification of the complete restabilization.

This free-Shercliff-layer instability exhibits strong similarities to
the expected behavior of the standard MRI in a Taylor-Couette device
because in both cases a magnetic field acts to destabilize otherwise
stable flow and in both cases the associated angular momentum
transport results in a large modification to the azimuthal velocity
profile. But this instability is a hydrodynamic instability on a
background state established by the magnetic field and is present with
$Rm \ll 1$. While there are inductionless relatives of the standard
MRI, such as the so-called HMRI which relies on azimuthal and axial
applied magnetic fields~\cite{hollerbach.prl.2005, stefani.njp.2007},
the unimportance of induction here is in stark contrast to the
requirement of a finite minimum $Rm$ for the standard MRI in an axial
magnetic field.

These results have particular relevance to other MHD experiments in
which similar shear layers may be established. A spherical Couette MHD
experiment produced a nonaxisymmetric instability with applied
magnetic field that was claimed to be
 the
MRI~\cite{sisan.prl.2004}. However, subsequent simulations have
attributed those observations to hydrodynamic instability of free
shear layers~\cite{hollerbach.prsa.2009, gissinger.pre.2011}, similar
to the observations that we report. We expect that other cylindrical
devices, such as the PROMISE 2 experiment~\cite{stefani.pre.2009},
could produce this instability. But the critical value of $\Lambda$
will likely change for experiments with different geometric aspect
ratios.

The free-Shercliff-layer instability is not expected to impact the
study of the MRI in this device since the magnetic fields required for
the MRI are weaker than those required for the Shercliff layer
instability at MRI-relevant speeds~\cite{gissinger.pof.2012}.

This work was supported by the U.S. Department of Energy's Office of
Sciences - Fusion Energy Sciences Program under contract number
DE-AC02-09CH11466, the U.S. National Science Foundation under grant
numbers AST-0607472 and PHY-0821899, and the U.S. National Aeronautics
and Space Administration (NASA) under Grant No. APRA08-0066
and No. ATP06-35.

\bibliography{thebibliography}

\end{document}